\documentstyle[12pt]{article}

\title{C-Periodicity and the Physical Mass
        in the 3-State Potts Model}

\author{R. V. Gavai{$^1$} \\
{\normalsize International Centre for Theoretical Physics, Trieste, Italy
} \\[3mm]
{\normalsize and} \\[3mm]
L. Polley\\
{\normalsize FB Physik, Oldenburg University, 26111 Oldenburg, Germany
 } }

\begin{document}

\maketitle

\begin{abstract}

The standard infinite-volume definition of connected correlation function and
particle mass in the 3-state Potts model can be implemented in Monte Carlo
simulations by using C-periodic spatial boundary conditions. This avoids
both the breaking of translation invariance (cold wall b.c.) and the
phase-dependent and thus possibly biased evaluation of data (periodic b.c.).
The numerical feasibility of the standard definitions is demonstrated by
sample computations on a $24\times24\times48$ lattice.

\end{abstract}

\footnote{Permanent address: Tata Institute of Fundamantal Research,
Bombay 400 005, India}

\newpage

The 3-dimensional 3-state Potts model has been extensively studied
in particle physics context because of its relation to the
finite-temperature SU(3) Yang-Mills theory \cite{Yaffe}.
Both have phase transitions which are
characterised by spontaneous breaking of a global $Z(3)$-symmetry.
A controversial issue for the latter has been the order of the
deconfinement phase transition\cite{Bacilieri,Brown}
While two coexisting states, typical of a first order phase transition,
were reported in Ref. \cite{Brown}, long-range correlations reminiscent
of a second order phase transition were observed in Ref. \cite{Bacilieri}.
The resolution of this controversy was first proposed in the context of
the Potts model in Ref. \cite{GaKaPe}.  Using finite size scaling
theory for bulk quantities, a first order phase transition was established
although the correlation length displayed identical behaviour to that
seen in Ref. \cite{Bacilieri}.  It was argued that the diverging
correlation length was not due to a massless physical particle,
but rather due to the finite-size effect of
vacuum tunnelling.  Subsequently, simulations of $SU(3)$
gauge theory at finite temperature confirmed \cite{Ukawa}
such an interpretation.  The problem of determination of the physical
correlation length was, however, left almost unsolved.  Since
the physical picture of a phase transition appears intuitively
much clearer in terms of the physical correlation length, it
appears desirable to devise a method of separating it from the
tunnelling correlation length.  Indeed, various physics applications,
such as the Early Universe or heavy ion collsions, need a precise value
of the physical correlation length in the critical region.

Most simulations employ periodic spatial boundary conditions to
minimise surface effects.  These, however, give rise to
a rather complicated low-energy level structure in the broken phase
of a theory with a spontaneously broken discrete symmetry.  For both
the Potts model and the finite temperature $SU(3)$ theory, one has
three ordered vacua and tunnelling occurs between them in a finite
volume. Close to the transition temperature there is order-disorder
tunnelling as well.  Thus, if the mass of the physical particle were
small, it would be hard to separate it from the multitude of tunnelling
levels.  Indeed, various prescriptions \cite{GaKaPe,Fukugita,Gavai}
have been used to extract the physical mass in the Potts model.
These are, however, not satisfactory since the results can be
sensitive to various assumptions and parameters.  Thus, e.g.,
Ref.\,\cite{Fukugita} claimed a discontinuity in the physical mass which
was shown to decrease with increasing volume in Ref.\,\cite{Gavai}.
We therefore agree with the authors of Ref.\,\cite{Bacilieri} that a useful
strategy is to single out one of the three ordered ground states by
imposing a suitable boundary condition, thus simplifying the low-energy
level structure.
Our boundary condition will emulate spatial infinity in two respects:
breaking the global Z(3) symmetry (without an external field) while
preserving translational invariance.

A Z(3)-symmetry breaking spatial boundary condition already extensively
studied for SU(3)
Yang-Mills theory is the cold wall \cite{Bacilieri}: all links lying
{\em in} the wall are set
to their trivial values. The Potts model analogue is to put $s_{\rm w}=1$ for
all spins on the wall.  The cold wall, however, breaks the translational
invariance.  In order to reduce surface effects, one is forced to obtain
results away from the wall where one sees little difference from the
periodic boundary conditions.

The virtue of periodic b.c., on the other hand, is their preservation of
translational invariance, and avoidance of any surface interactions. With a
cold wall b.c.\ the very concept of particle mass is only an approximation.
Furthermore, particles with any kind of charge will induce anticharges on the
wall and will thus incur surface interactions.

Recently, C-periodic spatial boundary conditions have been
developed for Monte Carlo simulations  of (generalized)
non-zero charges on finite lattices \cite{Polley,Kronfeld,Wiese}.
A numerical study of the
SU(3) Polyakov loop using those b.c.\ has already been carried out
\cite{Wiese}.
In this Letter, we propose C-periodic spatial boundary conditions for
extracting the physical mass of the 3-dimensional 3-state Potts model in
Monte Carlo simulations.
As translational invariance is preserved, the concept of physical mass is
clearly defined, and a precise determination by standard methods of statistical
mechanics is possible. Connected correlation functions can be defined without
any ad-hoc prescriptions (such as phase separation or arbitrary subtraction).

The Hamiltonian of the Potts model with nearest-neighbour couplings is
$$
H = -  \sum_{\langle j,k\rangle} \delta_{\sigma_j,\sigma_k}
$$
where $\sigma_i\in\{0,1,2\}$ defines the spin on site $i$.
The partition function is
$$
Z = \sum_{\{\sigma\}} e^{- \beta H}
$$
Link by link, $H$ is invariant under the charge-conjugation operation
$$
s \rightarrow \bar{s} \quad\mbox{where}\quad s = e^{2\pi i \sigma / 3}
                      \quad\mbox{and}\quad \bar{s} = e^{-2\pi i \sigma / 3}.
$$
In terms of the $\sigma$s,
$$
0\stackrel{\rm C}{\rightarrow}0 \qquad
1\stackrel{\rm C}{\rightarrow}2 \qquad
2\stackrel{\rm C}{\rightarrow}1 .
$$
We consider C-periodic spin configurations on an $L^3$ lattice,
\begin{equation}
\sigma_{i+L} = \bar{\sigma}_i~~~,~~~\label{cbc}
\end{equation}
where $i+L$ denotes the site $i$ translated by the lattice length $L$ in a
spatial
direction. Such configurations are conveniently visualized as a chess-board
array of copies and
C-conjugate copies of the $L^3$ configuration stored in the computer. Under
translation
$\sigma_i \rightarrow \sigma_{i+1}$
of a C-periodic $\sigma$ configuration the Hamiltonian is invariant.

The ground state of the ordered phase
is characterized by $\sigma_i=0$ on all sites $i$ and is C-invariant.
It is the same vacuum as with periodic boundary conditions and has the same
energy.
There are no other C-periodic vacua: if $\sigma_i = 1$ for some $i$ then
$\sigma_j = 2$
must also occur, so the minimum excitation energy is that of six $\langle
ij \rangle$ bonds
which corresponds to a {\em particle} excitation. A configuration with
all $\sigma_i=1$ on the $L^3$ lattice will have all $\sigma_j=2$ on the
adjacent C-conjugate
lattices and will have the excitation energy of $n_{\rm cbc}L^2$ bonds
($n_{\rm cbc}=$ number of C-periodic directions). It is a many-particle state
rather than another vacuum.

The infinite-volume definition of the connected spin-spin correlation function
$$
C(x-y) = \langle \bar{s}_x s_y \rangle - \langle \bar{s}_x \rangle \langle s_y
\rangle
$$
makes sense with C-periodic boundary conditions, too, since translation
invariance is
preserved. Due to C-conjugation invariance of the Hamiltonian, the
correlation function is real. Furthermore,
\begin{eqnarray*}
      C      &   =    &   R + I   \\
\mbox{where} & R(x-y) & = \langle {\rm Re}\,s_x {\rm Re}\,s_y \rangle
                      - \langle {\rm Re}\,s_x \rangle \langle {\rm Re}\,s_y
                      \rangle \\
\mbox{and}   & I(x-y) & = \langle {\rm Im}\,s_x {\rm Im}\,s_y \rangle
\end{eqnarray*}
In any coordinate direction with a C-periodic boundary condition, Re$\,s$ is
periodic while Im$\,s$ is {\em antiperiodic} \cite{Polley,Wiese}. Thus
the $R$ and $I$ parts of the correlation function $C$ are clearly
distinguished here. By contrast, both $R$ and $I$ would be periodic
functions if purely periodic b.c.\ were imposed.

The Fourier decomposition of $C(x-y)$ can be interpreted in terms of
``particle''
excitations of the two-dimensional quantum field theory underlying the
 three-dimensional
Potts model. This requires that a direction of Euclidean time is singled out
in which the
boundary condition is periodic (finite temperature). Of the two remaining
spatial
directions, at least one should be C-periodic for lifting the degeneracy of
the ordered vacua.

For particles generated by Re$\,s$ the projection onto zero spatial momentum
can be
implemented as usual with periodic b.c. The masses of such particles are
best obtained
from the correlation function of Re$\,s$ averaged over the two-dimensional
time slices.

For particles generated by Im$\,s$, there is a minimal spatial momentum of
$\pi/L$ for
each C-periodic direction \cite{Polley}. Hence, averaging Im$\,s$ over a time
slice would
mean to project onto an {\em antiperiodic continuation} of the constant
function--- an oscillatory step
function with a strong contamination of higher momenta. The closest
approximation to zero momentum in the antiperiodic case is a projection
onto functions
of minimal spatial momentum: $\cos x\frac{\pi}{L}$ and $\sin x\frac{\pi}{L}$
for C-periodicity in the x direction, and analogously for the y direction.
As a consequence, the mass $m_-$ of such a particle is not directly
observable in the Im $s$ correlation function;
rather, its correlation length is given by the energy $E$ which depends
on the spatial momentum according to some dispersion
relation. Exploiting the cubic symmetry of the Hamiltonian, which
includes space inversion and rotations by $\pi/2$
about the coordinate axes, one finds that an inverse propagator
for the Potts model must take the form
     $ m^2 + p_1^2 + p_2^2 + p_3^2 + {\cal O} (p^4) $.
Hence, identifying $p_3=iE$, the exponential decay of the correlation
function is related to the particle mass and the spatial momentum by
\begin{equation} \label{Emin}
 E^2 = m^2 + p_1^2 + p_2^2 + {\cal O} (p^4,p^2 E^2,E^4)
\end{equation}
In this Letter, eq.\,(\ref{Emin}) will be used only at minimal momentum:
$p_{i~{\rm min}} = 0$ or $\pi/24$, depending on whether a periodic or
C-periodic boundary condition is imposed in the corresponding direction.
Furthermore, the energy levels we consider will be of order 0.1
in lattice units. Thus ${\cal O}(p^4,p^2 E^2,E^4)$ amounts to a
correction in the 1\% range. This is unresolvable within our statistical
errors and will be neglected.

Since the C-periodic boundary conditions allow only one ordered state for
$\beta > \beta_c$, vacuum tunnelling can take place
only in the coexistence region of width $\sigma_L$ around
$\beta_{c,L}$ on a finite $L^3$ lattice.  As a consequence, possible
contamination of physical mass by the tunnelling mass is restricted to
this region only.  Figure \ref{Schem} shows schematically the expected
behaviour of the lowest excitation energy, near the phase transition,
as a function of $\beta$.  Also shown is the corresponding behaviour for
the periodic boundary case \cite{GaKaPe}.

Following Wang and DeTar \cite{WDT92}, we estimate the tunnelling
effects for our version of the Potts model in terms of a simple model.
Compared to their case,
the Potts model with C-periodic boundary conditions has a much reduced
symmetry---only charge conjugation symmetry is left.
We, therefore, choose an
extensive ``ordered'' state $|{\rm O}\rangle$ which derives from the
exact ordered vacuum
at infinite $\beta$, and another extensive state $|{\rm D}\rangle$ which
derives from the exact disordered vacuum at $\beta=0$
to descibe our model.  Both these states
are assumed to be invariant under charge conjugation.
In case of $|{\rm O}\rangle$ this is because all spins $s$ are equal to
the real number 1;
in case of $|{\rm D}\rangle$ it is due to the superposition of all spin
eigenstates with equal amplitudes. We furthermore assume the following
relations to hold for the matrix elements of the real part $\hat{r}$ of
the spin operator:
$$
   \langle {\rm O} | \hat{r} | {\rm O} \rangle = 1 \qquad
   \langle {\rm O} | \hat{r} | {\rm D} \rangle =
   \langle {\rm D} | \hat{r} | {\rm D} \rangle = 0
$$
When restricted to the Hilbert space spanned by
$|{\rm O}\rangle$ and $|{\rm D}\rangle$ the Hamiltonian will take the
form
$$ \left( \begin{array}{cc} E_{\rm O} & c \\ c & E_{\rm D}
                                                \end{array} \right) $$
where $c\geq0$ as a matter of phase convention for the two states.
At the point of order-disorder coexistence the energies $E_{\rm O}$ and
$E_{\rm D}$ are equal. More generally, they depend on $\beta$ and are
proportional to the spatial volume $V$ except for boundary effects.
The matrix element $c$ describes the vacuum tunnelling to be expected
for $V<\infty$; it will depend on $\beta$ and will decrease with
$V\rightarrow\infty$.
For $c\neq0$ the eigenstates of the Hamiltonian are superpositions
of the form
\begin{eqnarray*}
 |+\rangle &=& \sin\alpha |{\rm O}\rangle + \cos\alpha |{\rm D}\rangle\\
 |-\rangle &=& \cos\alpha |{\rm O}\rangle - \sin\alpha |{\rm D}\rangle
\end{eqnarray*}
\begin{equation} \label{alpha}
\mbox{where}\quad  \tan\alpha = \sqrt{1+x^2} + x \quad\mbox{ with }\quad
                               x = \frac{E_{\rm O}-E_{\rm D}}{2c}
\end{equation}
Hence we have
$$
   \langle + | \hat{r} | + \rangle = \cos^2\alpha \qquad
   \langle - | \hat{r} | - \rangle = \sin^2\alpha \qquad
   \langle + | \hat{r} | - \rangle = \langle - | \hat{r} | + \rangle =
                                                 \sin\alpha \cos\alpha
$$
The energy eigenvalues are
$$
E_{\pm} = \frac{E_{\rm O}+E_{\rm D}}{2} \pm \frac{\epsilon}{2}
\quad\mbox{ where }\quad
\epsilon = \sqrt{(E_{\rm O}-E_{\rm D})^2 + 4 c^2 }
$$
Let $| - \rangle$ correspond to the lower energy level which we put to
zero, and let $| + \rangle$ be the first excited state at
energy $\epsilon$. The partition function
is $$ Z = 1 + e^{-\beta\epsilon} $$
Evaluating traces in the $|\pm\rangle$ basis we find for the
statistical expectation value of the real part of the spin
$$
\langle \hat{r} \rangle = \frac{\cos^2\alpha +
                                         e^{-\beta\epsilon}\sin^2\alpha}
                               { 1 + e^{-\beta\epsilon} }
$$
To recover the canonical picture, let us assume that the spatial volume
is finite but large. At the coexistence point, $\epsilon\approx 0$ so that
$ \langle \hat{r} \rangle \approx \frac12 $.
Elsewhere, $\epsilon\propto V$, hence the exponential factors are very
small, and $ \langle \hat{r} \rangle \approx \cos^2\alpha $. Furthermore,
eq.\,(\ref{alpha}) implies $\alpha\approx 0$ for $E_{\rm O} < E_{\rm D}$
(broken phase) and $\alpha\approx\frac{\pi}2$ for $E_{\rm O} > E_{\rm D}$
(symmetric phase). Thus the spin expectation value indeed meets our
expectations in this vacuum model. As for the correlation function,
$$
\langle \hat{r}(0) \hat{r}(\tau) \rangle = Z^{-1}~{\rm Tr} \left( \hat{r}
                       e^{-\tau H} \hat{r} e^{\tau H} e^{-\beta H} \right)
$$
we have
\begin{equation}
\langle \hat{r}(0) \hat{r}(\tau) \rangle =
\frac{\sin^2\alpha\cos^2\alpha}{\cosh(\beta\epsilon/2)}
           \cosh \epsilon(\tau - {\textstyle\frac12}\beta)
+
\frac{\cos^4\alpha + \sin^4\alpha e^{-\beta\epsilon}}{1+e^{-\beta\epsilon}}
{}~~.~~ \label{rort}
\end{equation}
Right at the phase transition where $\cos\alpha=\sin\alpha
=\sqrt{1/2}$ this expression simplifies to
$$
\langle \hat{r}(0) \hat{r}(\tau) \rangle = \frac14
              \frac{\cosh \epsilon(\tau - \frac12 \beta)}
                   {\cosh(\beta\epsilon/2)}~ +~ \langle\hat{r}\rangle^2
$$
As for particle excitations, we assume that one-particle states
are generated by acting with the local spin operator $\hat{s}(x)$ on a
vacuum state.
The real part $\hat{r}(x)$ of the spin field operator has C-parity $+1$
while the imaginary part $\hat{i}(x)$ has C-parity $-1$. The Hamiltonian
$H$ and the states $|{\rm O}\rangle$ and $|{\rm D}\rangle$
are charge-conjugation invariant. In the spin-spin correlations we thus
expect to see an excitation scheme whose levels can be classified by
their C-parity.

These model considerations suggest that vacuum tunnelling in the C-periodic
Potts model (order-disorder tunnelling) only occurs in a narrow region
around the phase transition, as illustrated in Figure \ref{Schem}.
Outside that region, both in the ordered and disordered
phase, the energy spectrum consists of a
nondegenerate extensive ground state, followed by particle excitations.
In the simplified model described above, $\epsilon$ is the energy of the
first excited state which corresponds to the physical mass sufficiently
away from $\beta_c$.  In the coexistence region, however, $\epsilon
\approx c$ and is thus governed entirely by the finite volume effects.
Hence on increasingly larger volumes $\epsilon$ becomes vanishingly small at
the transition point. On the other hand, the width of the coexistence region
also goes down linearly with volume, thus opening the possibility of
obtaining the physical mass at $\beta_c$ in a limiting procedure.

We have simulated the 3-dimensional 3-state Potts model on a
$24\times24\times48$ lattice, imposing C-periodic boundary
conditions in {\em one} of the spatial ($L=24$) directions,
at couplings $\beta=0.5505$, $\beta=0.551$ and $\beta=0.55125$.
A total of 40000 measurements were performed at each coupling.
At $\beta=0.5505$ and $\beta=0.551$ measurements were interspersed with
20 updatings of the spin configuration; at $\beta=0.55125$ with 100 updatings.
Plots of the MC results for those couplings are very similar.
We here present plots only for $\beta=0.5505$.

Figure \ref{History} shows the Monte Carlo history of the real part of the
spin averaged over the 3-dimensional lattice. The Figure illustrates the
tunnelling between a globally ordered and a globally disordered state.

Our MC data for the Re$\,s$ two-point
correlation function $R(d)$, projected onto zero spatial momentum
(Figure \ref{RUnsub}) are strongly dominated by a constant---presumably
the one modelled by eq.\,(\ref{rort}).
The measurement-to-measurement fluctuations in $R(d)$, also, are similar for
all values of $d$, supporting the interpretation as a fluctuating constant.
In fact, the errors in the exponentially decaying particle
contributions which we are interested in are much smaller. To extract
the latter, a better observable, suggested by eq.\,(\ref{rort}), is
the differences $R(d)-R(d+1)$. The constant part of the spin correlation
thus drops out from both the mean values {\em and} the error bars, as is
evident from Figure \ref{Rdiff}. A one-particle contribution
to this difference would be of the form
$A \sinh m (\frac12 N_{\beta}-\frac12-d)$ where $N_{\beta}$ is the temporal
lattice size. Figure \ref{Rdiff} shows the corresponding fit of the data.
The mass of the excitation is given in the Table.

Practically the same mass is obtained by a global $\chi^2$ fit
of the form $A\cosh m(N_{\beta}-d)+B$ to the unsubtracted function $R(d)$.
In this latter way one obtains an excessively shallow $\chi^2$ minimum
since $\chi^2$ is then mainly based on the fluctuations of the constant.
Nevertheless, the agreement of mean excitation energies obtained from our
$R(d)$ and from $R(d)-R(d+1)$ supports the conjecture that C-periodic
boundary conditions eliminate order-order tunnelling so that vacuum tunnelling
effects must be taken into account only near the phase transition
(order-disorder tunnelling).

For the correlation of the imaginary part of the spin, projected onto
minimal spatial momentum, Figure \ref{Idiff} shows the differences
$I(d)-I(d+1)$ and a fit of the form $A \sinh E (\frac12 N_{\beta}-\frac12-d)$.
The mass of the C$=-1$ particle as obtained from formula
(\ref{Emin}) is given in the Table. In view of the non-zero momenta involved,
we expect to find, as a consistency check, {\em no} particle energies in the
Im$\,s$ correlation \\ \hspace*{5mm}
(a) below $\pi/L$ if we impose C-periodicity in only one spatial direction,
\\ \hspace*{5mm}
(b) below $\sqrt{2\pi^2/L^2}$ if both spatial directions are C-periodic.\\
In our sample computations of case (a), $L=24$ so that the physicality bound is
0.131. This bound is consistent with the $\chi^2$ distributions we obtain at
all couplings from the {\em differences} $I(d)-I(d+1)$. In the unsubtracted
correlation functions $I(d)$ we actually observe small {\em constant}
contributions as well.
Furthermore, at all three couplings we find error bars consistent with
the fluctuations of the mean values {\em only} in $I(d)-I(d+1)$ while for
$I(d)$ the mean values are overlaid again with the error bars of a
fluctuating constant. However, the size of the effect is such that it
can easily be understood in terms of mere statistical fluctuations.
The constant fitted to $I(d)$ at $\beta=0.5505$, for example,
is $1.2\times10^{-4}$ which is less than a standard deviation, after
40000 measurements, of a fluctuating constant with zero average and
an {\sc rms} of the same order as in $R(d)$.
In the following, we shall form differences $I(d)-I(d+1)$ with the
intention to eliminate from $I(d)$ the artifacts of a finite number of
iterations.

At $\beta=0.551$ and $\beta=0.55125$, where we expect to be sufficiently deep
in the ordered phase, the maximum-likelihood estimates (mean values in Table 1)
of $m_- = 0.11$ for both cases indicate a massive particle in the $C=-1$
channel degenerate with the $C=+1$ particle discussed above.
Since $m^2_-$ according to formula (\ref{Emin}) involves a difference,
$E^2 - p^2_{i~{\rm min}}$, the mass estimates in the
$C=-1$ channel can be quite sensitive to uncertainties in the correlation
length $\xi=E^{-1}$. At the above couplings this results in a rather large
error estimate for $m_-$.
At $\beta=0.5505$, i.e.\ near the phase transition, we find an essentially
unchanged $m_- = 0.13\pm0.02$, while $m_+$ decreases to 0.07 indicating
the increasing dominance of order-disorder tunnelling in this regime.

In conclusion, we have shown that the C-periodic boundary conditions, given
by eq.\,(\ref{cbc}), eliminate order-order tunnellings and allow numerical
implementation of the standard statistical-mechanical definition of
connected correlation functions and particle masses in the 3-dimensional
3-state Potts model.  This is also true for all $Z(N)$ models,
$N$-state Potts models and $SU(N)$ gauge theories at finite temperature
in any dimensions, provided $N$ is odd.  In particular, quenched QCD
also falls in this category.  Furthermore, we suggest a method to obtain
the physical mass in all these theories at $\beta_c$ in the thermodynamic
limit.  It will be interesting to find out using this method
whether the first order phase
transition in the 3-dimensional 3-state Potts model or the 3-dimensional
$SU(3)$ gauge theory at nonzero temperature manifests itself as
a discontinuity in the physical correlation length.

Also, it will be interesting to analyse the dispersion relation
(\ref{Emin}) beyond the quadratic order in the energy and momenta.
For quasi-free actions with a Laplacian
structure of derivatives,
$E^2=m^2+\sum_k \sin^2 p_k$ is a natural extension; it has,
in fact, produced satisfactory mass corrections for U(1) monopoles
in the Villain approximation on smaller lattices \cite{Polley}.
For the Potts model, the action is of an entirely different form.
Here one can study numerically the correlation functions of the real
and imaginary parts of the spin after projection
onto various non-minimal spatial momenta. However, on lattices large
enough so that ${\cal O}(p^4_{\rm min})$ is negligible, the Im\,$s$
correlation function in conjunction with dispersion relation
(\ref{Emin}) offers perhaps the best possible way to obtain the
mass of the lowest $C =-1$ particle as it is not overlaid with any
vacuum tunnelling.

\subsection*{Acknowledgments}

R.V.G. wishes to thank the computer centre staff of ICTP, Trieste,
for providing time on their CONVEX-210 and their kind help.
Thanks are also due to Professor Abdus Salam, the International Atomic
Energy Agency, and UNESCO for hospitality at ICTP.
L.P. wishes to thank Dr.E.Mendel for a discussion of the statistical
errors, and Landesrechenzentrum f\"ur die Wissenschaft at Bremen for
providing CPU time on their VP200.

\newpage

\begin{figure}
\caption{Schematic diagram illustrating the behaviour of the energy
of the first excited state as a function of $\beta$ near the phase
transition for two volumes $V_1 < V_2$, using C-periodic
(full lines) and periodic (full and dashed lines) boundary
conditions. \label{Schem}}
\vspace{1cm}
\end{figure}

\begin{figure}
\caption{Monte Carlo history of the real part of the spin
(3-dimensional average) at $\beta=0.5505$.\label{History}}
\vspace{1cm}
\end{figure}

\begin{figure}
\caption{Monte Carlo data for the Re$\,s$ correlation $R(d)$
at $\beta=0.5505$ and fit of the form $A\cosh m(24-d)+B$.
The error bars are dominated by the fluctuations of the constant $B$.
\label{RUnsub}}
\vspace{1cm}
\end{figure}

\begin{figure}
\caption{Monte Carlo data for the differences $R(d)-R(d+1)$
at $\beta=0.5505$ and fit of the form $A\sinh m(23.5-d)$.
The $\chi^2$ fit is based on data points $5\leq d \leq 23$.\label{Rdiff}}
\vspace{1cm}
\end{figure}

\begin{figure}
\caption{Monte Carlo data for the differences $I(d)-I(d+1)$
at $\beta=0.5505$ and fit of the form $A\sinh E (23.5-d)$.
The $\chi^2$ fit is based on data points $5\leq d \leq 23$.\label{Idiff}}
\vspace{1cm}
\end{figure}

\newpage

\begin{table}
\caption{Inverse correlation lenghts $\xi^{-1}$ and particle
masses obtained from the spin-spin correlation. $R$ and $I$
refer to the correlation of the real or imaginary parts of the
spin, respectively. The form of the fit is
$ A\sinh\xi^{-1}(23.5-d)$.}
\vspace{2mm}
\begin{center}
\begin{tabular}{|c|c|c|c|} \hline
$\beta$ &   Function   &  $\xi^{-1}$ & Particle Mass \\
\hline
0.55050 & $ R(d)-R(d+1) $
& $ 0.072
\begin{array}{c} + \, 0.012 \\ - \, 0.014 \end{array}
$
& same as $\xi^{-1}$
\\ \hline
0.55050 & $ I(d)-I(d+1) $ &
$ 0.184 \begin{array}{c} +\,0.016 \\ -\,0.014 \end{array} $ &
$ 0.13 \pm 0.02 $
\\
\hline
0.55100 & $ R(d)-R(d+1) $
& $ 0.108
\begin{array}{c} + \, 0.012 \\ - \, 0.014 \end{array}
$
& same as $\xi^{-1}$
\\ \hline
0.55100 & $ I(d)-I(d+1) $ &
$ 0.17 \begin{array}{c} +\,0.05 \\ -\,0.045 \end{array} $ &
$ 0.11 \begin{array}{c} +\,0.07 \\ -\,0.11 \end{array} $
\\
\hline
0.55125 & $ R(d)-R(d+1) $ &
$ 0.134
\begin{array}{c} +\,0.008 \\ -\,0.007 \end{array} $ &
same as $\xi^{-1}$
\\ \hline
0.55125 & $ I(d)-I(d+1) $ &
$ 0.17
\begin{array}{c} +\,0.045 \\ -\,0.055 \end{array} $ &
$ 0.11
\begin{array}{c} +\,0.06 \\ -\,0.11 \end{array} $
\\ \hline
\end{tabular}
\end{center}
\end{table}

\end{document}